\documentclass[%
 reprint,
nofootinbib,
 amsmath,amssymb,
 aps,
]{revtex4-2}

\usepackage{graphicx}
\usepackage{dcolumn}
\usepackage{bm}
\usepackage[colorlinks=true,citecolor=blue,urlcolor=blue,linkcolor=red]{hyperref}
\usepackage{physics}
\usepackage{siunitx}
\usepackage{pgf}
\usepackage[normalem]{ulem}
\usepackage{xspace}
\usepackage{multirow}

\newcommand\Xe{X_{\rm e}}
\newcommand\rs{r_s^{\rm d}}

\newcommand{\mnras}{Mon. Not. Roy. Astron. Soc.}
\newcommand{\apjl}{Astrophys. J. Lett.}

\newcommand{\aap}{Astron. Astrophys.}
\newcommand{\aapr}{Astron. Astrophys. Rev.}
\newcommand{\pasp}{Publ. Astron. Soc. Pac.}
\newcommand{\jcap}{JCAP}
\DeclareUnicodeCharacter{2212}{-}

\newcommand{\planck}{{\it Planck}}
\newcommand{\lcdm}{\mbox{$\Lambda$CDM}}
\newcommand{\modrec}{\texttt{ModRec}}
\newcommand{\sptthreeg}{SPT-3G}
\newcommand{\shoes}{SH0ES\xspace}

\newcommand{\comment}[1]{{}}

\makeatletter
\newcommand\thefontsize[1]{{#1 The current font size is: \f@size pt\par}}
\makeatother

\begin{document}

\preprint{APS/123-QED}

\title{DESI and the Hubble tension in light of modified recombination}

\author{Gabriel P. Lynch$^1$ }
\email{gplynch@ucdavis.edu}
\author{Lloyd Knox$^1$}
\author{Jens Chluba$^2$}

\affiliation{
$^1$Department of Physics and Astronomy, University of California, Davis, CA, USA 95616
\\
$^2$Jodrell Bank Centre for Astrophysics, University of Manchester, Manchester M13 9PL, UK
}

\date{\today}

\begin{abstract}
Recent measurements and analyses from the Dark Energy Spectroscopic Instrument (DESI) Collaboration and supernova surveys combined with cosmic microwave background (CMB) observations, indicate that the dark energy density changes over time. Here we explore the possibility that the dark energy density is constant, but that the cosmological recombination history differs substantially from that in \lcdm. When we free up the ionization history, but otherwise assume the standard cosmological model, we find the combination of CMB and DESI data prefer i) early recombination qualitatively similar to models with small-scale clumping, ii) a value of $H_0$ consistent with the estimate from the \shoes Collaboration at the $2\sigma$ level, and iii) a higher CMB lensing power, which takes pressure off of otherwise tight constraints on the sum of neutrino masses. Our work provides additional motivation for finding physical models that lead to the small-scale clumping that can theoretically explain the ionization history preferred by DESI and CMB data. 

\end{abstract}

\maketitle


\section{\label{sec:intro}Introduction}

The standard \lcdm\ cosmological model continues to provide an excellent fit to measurements of the cosmic microwave background (CMB) \citep{Planck:2018vyg, SPT-3G:2022hvq, ACT:2020gnv, ACT:2023kun}, baryon acoustic oscillations (BAO) \citep{BOSS:2016wmc, eBOSS:2020yzd, DESI:2024mwx}, and uncalibrated type Ia supernovae (SNe) \citep{Brout:2022vxf, Rubin:2023ovl, DES:2024tys}, while the Hubble constant tension \citep[e.g.][]{Verde:2023lmm} persists as a potential indication of new physics beyond \lcdm\ \citep{Valentino2021H0, Schoneberg:2021qvd}. More recently, BAO measurements from the Dark Energy Spectroscopic Instrument (DESI), in combination with uncalibrated SNe measurements, suggest evolving dark energy, with a significance ranging from $2.5\sigma$ to $3.9\sigma$ depending on the supernova dataset used \cite{DESI:2024mwx}.

The cosmological recombination model plays an important role in interpreting early and late-time cosmological probes \citep[e.g.,][]{Hu1995, Chluba2005, Lewis2006}: it determines the sound horizon $\rs$, which is a standard ruler used to obtain $H_0$ from CMB data and cosmological distances from BAO data. In this paper we explore the cosmological implications of relaxing the strong theoretical prior on recombination that is implicit in the \lcdm\ model. Using the model-independent framework recently developed in \cite{Lynch:2024gmp}, constrained by the combination of \planck\ and DESI BAO data, we preserve a cosmological constant\footnote{By construction our model is \lcdm\ except for the ionization history during recombination}, reduce the tension with the  \shoes \citep{Riess:2021jrx} $H_0$ measurement to $2 \sigma$, and potentially solve the cosmological neutrino mass problem articulated in \cite{Craig:2024tky}. On the other hand, we also see a increased tension with the $\Omega_{\rm m}$ value inferred from several supernova datasets. Although we currently have no explanation for this tension, the increase compared to \lcdm\ is only at the 0.4 to 0.5$\sigma$ level, depending on SNe dataset used in the analysis.

The preference for a high $H_0$ is a new development in the role played by BAO data in understanding the Hubble tension. Previously, BAO measurements from the BOSS \citep{BOSS:2016wmc} and eBOSS \citep{eBOSS:2020yzd} collaborations favored distances $D_M(z)/\rs$ and $D_H(z)/\rs$ consistent with \planck\ when assuming \lcdm. While a high $H_0$ can be obtained from CMB data alone in models where the sound horizon $\rs$ is allowed to vary (such as modified recombination), this results in a change to low-redshift distances which are constrained by BAO measurements. As such, BAO data have posed a challenge for a number of solutions which address the Hubble tension through changes to recombination-era physics \citep[e.g.,][]{Hart2020H0, Sekiguchi:2020teg, Jedamzik2020Relieving, Thiele2021, Lee:2022gzh, Lynch:2024gmp}. It is possible to circumvent these constraints, for example by introducing a non-zero mean spatial curvature \citep{Sekiguchi:2020teg}, but with the recent DESI data this is no longer necessary. This change between datasets is almost entirely driven by the DESI luminous red galaxy (LRG) bin between $0.6 < z < 0.8$, which is itself in $3\sigma$ tension with measurements from SDSS \citep{DESI:2024uvr, DESI:2024mwx}.

The departures from the standard recombination scenario preferred by the phenomenological model used here contain features which are not common in physical models that modify recombination \citep[see also][]{Lynch:2024gmp}. However, we demonstrate that an appropriately smoothed version of the best-fit ionization history in our model fits the data almost as well, and bears qualitative similarities to scenarios in which recombination is altered due to small-scale clumping \citep{Rashkovetskyi:2021rwg, Galli:2021mxk, Jedamzik2020Relieving, Thiele2021}, featuring a highly inhomogeneous ionization fraction at very small scales. 

This finding is interesting on account of the growing number of indications that \lcdm\ may indeed hold surprises at small scales that could all be linked to the same physical origin. First, accelerated recombination due to small-scale density perturbations could be caused by primordial magnetic fields (PMFs) \citep[e.g.,][]{Jedamzik2020Relieving, Galli:2021mxk, Thiele2021}. Although the Hubble tension can be alleviated by a number of alternatives \citep[e.g.,][]{Schoneberg:2021qvd}, PMFs are naturally expected, even if their origin, evolution and overall properties are still not understood \citep[e.g.][]{Durrer2013, Subramanian2016, Vachaspati2021}.
Second, we know that every galaxy contains a supermassive black hole at its center, yet it is unclear if it is possible to form these without an intermediate mass black hole (IMBH) progenitor ($\simeq 10^3$ solar masses)  \citep[e.g.][]{Carr2021}. These IMBHs could be of primordial origin, stemming from enhanced small-scale curvature perturbations, and possibly even relate to the surprising rate of gravitational wave events \citep[e.g.,][]{Abbott2016, Bird2016} and the discovery of a stochastic GW background \citep[e.g.,][]{Agazie2023, Vaskonen2021}. 
Finally, the ARCADE radio excess \citep{2011ApJ...734....5F, 2018ApJ...858L...9D}, another persistent puzzle in cosmology \citep[see][for overview]{Singal:2017jlh,Singal2023, Cyr2023CSSol}, may be caused by an unresolved high-redshift radio source population, indicative of early structure formation. Similarly, the JWST observations of very high redshift galaxies \citep{Boylan-Kolchin:2022kae, 2024arXiv240518485C} could be another hint in this direction.
All these indications provide motivation for considering the possibility of a modified recombination epoch, and for reinterpreting the DESI data and Hubble tension in this context.

To be clear, the standard recombination scenario is well understood from a theoretical perspective \citep[e.g.,][]{Chluba:2010ca, 2011PhRvD..83d3513A}. Further, the CMB data themselves, which are highly sensitive to details of recombination, are consistent with the standard scenario. This is a non-trivial fact that should not be taken lightly, especially since the standard ionization history departs significantly from the equilibrium history. However, \citet{Lynch:2024gmp} found that a large range of ionization histories can also fit the data, and there has not yet been a more direct validation of the standard scenario by measurement of the cosmological recombination radiation \citep{Sunyaev2009, Chluba2016Cosmospec, Chluba2019Voyage, Hart2020Sensitivity}. Even without such measurements, forthcoming small-scale CMB temperature and polarization data from \sptthreeg\ stand to significantly improve constraints on this epoch \cite{Prabhu:2024qix, Lynch:2024gmp}, as will other CMB datasets to varying degrees, including ACT, SO \citep{2019JCAP...02..056A}, and CMB-S4 \citep{CMB-S4:2020lpa}. Our hypothesis is thus testable with data to come in the very near future. 

\citet{Pogosian:2024ykm} have recently performed an analysis with some similarities to our own. They adopt a \planck\ prior on $\Omega_{\rm m} h^2 \equiv \omega_{\rm m}$ and the angular size of the sound horizon, $\theta_{\rm s}$, and include these with DESI data while treating the comoving sound horizon as a free parameter. They find from the combination of DESI data and the \planck\ values of $\Omega_{\rm m} h^2$ and $\theta_{\rm s}$ (derived assuming \lcdm) that $H_0 = 69.88 \pm 0.93$ km/s/Mpc. Treating the comoving sound horizon as a free parameter greatly reduces the dependence of assumptions about recombination, as in our work. One practical difference between our approaches is that our framework allows for a data-driven reconstruction of the best-fitting ionization history. Additionally, we estimate $\Omega_{\rm m} h^2$ self-consistently, and find that the best-fit value rises above that of the best fit in \lcdm\ given \planck\ data, which is the $\Omega_{\rm m}h^2$ value used by \cite{Pogosian:2024ykm}.

In Section~\ref{sec:data}, we present parameter constraints using \planck\ and DESI data with a modified recombination epoch. In Section~\ref{sec:discussion}, we discuss the origin in the data of the high inferred value of $H_0$, as well as challenges to the modified recombination interpretation of DESI results. We also discuss features of the (smoothed) best-fit ionization history, demonstrating that apparent unphysical aspects do not matter much. We conclude in Section~\ref{sec:conclusion}.

\section{DESI data with modified recombination \label{sec:data}}

We use the phenomenological ``\modrec" model introduced in \citep[][henceforth LKC]{Lynch:2024gmp} to allow for added freedom in the recombination process, while going beyond the linear response approximation used in other related analyses \citep{Farhang:2011pt, Farhang:2012jz, Hart:2019gvj}. This framework introduces deviations, $\Delta \Xe(z)$, from the standard ionization history, $\Xe(z)^{\rm std}$, via 7 control points evenly spaced between $z=533$ and $z=1600$. The deviation is fixed to 0 at the boundaries of this interval. We note that this redshift range leaves both helium recombination as well as reionization unchanged. This parameterization is sufficient to approximate deviations predicted by physical models [see LKC] such as decaying or annihilating dark matter \citep{Adams:1998nr, Chen2004, Galli:2013dna, Slatyer:2016qyl, Finkbeiner:2011dx}, varying fundamental constants \citep{Planck:2014ylh, Hart:2017ndk, Hart2020H0, Sekiguchi:2020teg}, and PMFs \citep{Jedamzik2020Relieving, Galli:2021mxk}.

\begin{table*}
\caption{\label{tab:parameters}
Parameter values for the data combinations and models considered here. Percentages denote changes in the mean relative to the \lcdm\ \planck\ mean value. $A_s^\nu$ is the amplitude of the primordial power spectrum at $k=0.1/\text{Mpc}$, the scale where the ACT CMB data have the highest signal-to-noise on $\sum m_\nu$. The $\Delta \chi^2$ values are for the best fit to each model and data combination, and are relative to \lcdm\ using the same data combination.}
\begin{ruledtabular}
\fontsize{8pt}{8pt}\selectfont
\begin{tabular}{l c rl rl rl rl}

    & \multicolumn{3}{c}{\bfseries Planck} & \multicolumn{5}{c}{\bfseries Planck+DESI} \\ 
                \cline{2-4}                             \cline{5-10}
    & \lcdm & \multicolumn{2}{c}{\modrec} & \multicolumn{2}{c}{\lcdm} & \multicolumn{2}{c}{$w_0 w_a$CDM} & \multicolumn{2}{c}{\modrec}  \\
     $H_0$ [km/s/Mpc] & $67.36\pm 0.54$ & $69.4^{+6.7}_{-7.6}$ & $(\uparrow 3.0\%)$ & $68.10\pm 0.40$ & $(\uparrow 1.1\%)$ & $65.0^{+1.9}_{-3.4}$ & $(\downarrow 1.1\%)$ & $70.0\pm 1.1$ & $(\uparrow 3.9\%)$ \\
     $H_0$ tension & $4.9\sigma$       & \multicolumn{2}{c}{$0.5\sigma$} & \multicolumn{2}{c}{$4.4\sigma$}   & \multicolumn{2}{c}{$2.7\sigma$} & \multicolumn{2}{c}{$2.0\sigma$} \\
     $\Omega_{\rm m}$                     & $0.314\pm 0.007$& $0.305^{+0.041}_{-0.061}$ & $(\downarrow 2.7\%)$      & $0.304\pm 0.005$ & $(\downarrow 3.2\%)$  & $0.339^{+0.033}_{-0.024}$ & $(\uparrow 8.1\%)$ & $0.295\pm 0.007$ & $(\downarrow 5.9\%)$ \\
     $\Omega_{\rm m}$  tension  & $2.1\sigma$       & \multicolumn{2}{c}{$0.9\sigma$} & \multicolumn{2}{c}{$2.7 \sigma$}       & \multicolumn{2}{c}{$0.4\sigma$} & \multicolumn{2}{c}{$3.1\sigma$} \\
     $\Omega_{\rm m} h^2$                 & $0.142\pm 0.001$& $0.144\pm 0.005$ & $(\uparrow 1.0\%)$ & $0.141\pm 0.001$ & $(\downarrow 1.1\%)  $ & $0.142\pm 0.001$ & $(0.0\%)$ & $0.145\pm 0.002$ & $(\uparrow 1.6\%)$ \\
     $\rs h$ [Mpc]                & $99.07\pm 0.93$ & $101.0\pm 6.7$ & $(\uparrow 2.0\%)$ & $100.37\pm 0.69$ & $(\uparrow 1.3\%)  $ & $96.0^{+2.0}_{-4.4}$ & $(\downarrow 3.1\%)$ & $101.64\pm 0.97$ & $(\uparrow 2.6\%)$ \\
     $A_s \omega_{\rm m}^2$ [$\times 10^{11}$] & $4.253\pm 0.077$ & $4.32\pm 0.30$ & $(\uparrow 1.7\%)$ & $4.192\pm 0.071$ & $(\downarrow 1.4\%)$ & $4.237\pm 0.072$ & $(\downarrow 0.37\%)$ & $4.36\pm 0.13$ & $(\uparrow 2.6\%) $ \\
     $A_s^{\nu} \omega_{\rm m}^2$ [$\times 10^{11}$] & $4.151\pm 0.072$ & $4.21^{+0.28}_{-0.32}$ & $(\uparrow 1.4\%)$ & $4.104\pm 0.069$ & $(\downarrow 1.1\%)$ & $4.133\pm 0.067$ & $(\downarrow 0.43\%)$ & $4.24\pm 0.12$ & $(\uparrow 2.2\%) $ \\
     \hline
     $\Delta \chi^2_{\rm Planck}$   & 0.0               & \multicolumn{2}{c}{-7.2}        &  \multicolumn{2}{c}{0.0}               & \multicolumn{2}{c}{-2.8}        & \multicolumn{2}{c}{-5.5}      \\
     $\Delta \chi^2_{\rm DESI}$     & --                & \multicolumn{2}{c}{--}          &  \multicolumn{2}{c}{0.0}               & \multicolumn{2}{c}{-3.8}        & \multicolumn{2}{c}{-1.8}         \\
     $\Delta \chi^2_{\rm total}$    & 0.0               & \multicolumn{2}{c}{-7.2}        &  \multicolumn{2}{c}{0.0}               & \multicolumn{2}{c}{-6.6}        & \multicolumn{2}{c}{-7.2}          \\
\end{tabular}
\end{ruledtabular}
\end{table*}

\begin{table}
\caption{\label{tab:modrec_comparison} Cosmological parameter results for different data augmentations within the \modrec\ model. ``DESI$^{-}$" denotes the DESI dataset with the $z=0.71$ LRG data point removed. The ``DESI (smoothed)" column contains results using the full DESI data while fixing the recombination history to the smoothed best-fit variant in Fig.~\ref{fig:rec_history} and allowing cosmological parameters to vary, as discussed in Section~\ref{sec:discussion}. The final column contains constraints using SDSS data.}
\begin{ruledtabular}
\begin{tabular}{l c c c}
                   & \multicolumn{3}{c}{Planck + } \\
                                \cline{2-4}
                   & DESI$^{-}$         & DESI (smoothed)    & SDSS \\
$H_0$              & $69.1\pm 1.2$      & $71.05\pm 0.41$    & $68.9\pm 1.1$ \\
$H_0$ tension      & $2.5\sigma$        & $1.8 \sigma$       & $2.8\sigma$ \\
$\Omega_{\rm m}$         & $0.302\pm 0.008$ & $0.289\pm 0.005$     & $0.304\pm0.008$ \\
$\Omega_{\rm m}$ tension & $2.7\sigma$        &  $3.6\sigma$       & $2.7\sigma$ \\
\end{tabular}
\end{ruledtabular}
\end{table}

The control point parameters are included along with the six standard cosmological parameters $\vec{\theta} = \{\omega_{\rm b}, \omega_{\rm cdm}, H_0, n_s, \ln{10^{10} A_s}, \tau_{\rm reio}\}$ in an MCMC analysis using the \texttt{COBAYA} \citep{Torrado:2020dgo} sampler. For the theoretical calculations, we use an emulator created using the \texttt{CONNECT} \citep{Nygaard:2022wri} framework to accelerate the inference process. The emulator was trained to reproduce the output of a modified version of \texttt{CLASS} \cite{Blas:2011rf} which implements the \modrec\ model. Further details regarding this emulator are described in LKC. We use \planck\ 2018 TT+TE+EE and lensing data \citep{Planck:2019nip, Planck:2018lbu}, which we refer to as ``\planck". For BAO data, we use the recent DESI DR1 data \citep{DESI:2024uvr, DESI:2024lzq}. We also compare to the pre-DESI BAO measurements from BOSS/eBOSS \citep{BOSS:2016wmc,eBOSS:2020yzd} (including the SDSS Main Galaxy Sample \citep{Ross:2014qpa}), which we refer to collectively as ``SDSS".

Constraints on selected cosmological parameters in the \modrec\ model are presented in the final column of Table~\ref{tab:parameters}.
If recombination is allowed to vary from the standard model, \planck\ and DESI data together favor $H_0 = 70.0 \pm 1.07$ km/s/Mpc, a shift of $1.7 \sigma$ relative to the $H_0 = 68.1 \pm 0.4$ km/s/Mpc result if the standard recombination scenario is assumed. This is also higher than the preference of $H_0 = 68.9 \pm 1.1$ km/s/Mpc that is obtained using the pre-DESI BAO data from SDSS, which is shown in Table~\ref{tab:modrec_comparison}.

This development between datasets is strongly driven by the DESI $z=0.71$ BAO measurement using LRGs, which is itself in $\sim 3 \sigma$ tension with the SDSS measurements in similar redshift bins \citep{DESI:2024mwx}. The cause of this discrepancy is currently not understood, and the DESI collaboration suggests that it may be due to sample variance, in which case a reversion to nearer the SDSS value can be expected with more data. This is relevant to our analysis, as we find that if this data point is removed, the constraints using \planck+DESI are nearly the same as what is obtained using \planck+SDSS data (c.f. columns one and three of Table~\ref{tab:modrec_comparison}). 

\begin{figure}[t!]
    \centering
    \includegraphics{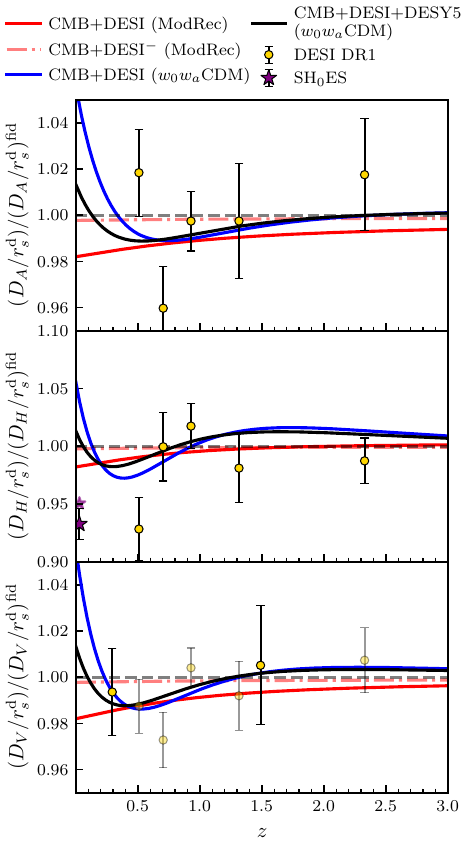}
    \caption{\textit{Top panel:} The transverse BAO scale, $D_A / \rs$ divided by a fiducial, which in this case is the best fit to \planck+DESI data assuming \lcdm. The red dot-dashed line shows the best fit in the \modrec\ model if the $z=0.71$ LRG data point is removed. We do not plot the $z=0.3$ and $z=1.49$ DESI data points, as $D_A$ and $D_H$ are not separately measured at these redshifts. \textit{Middle panel:} Similar to the top panel, but with $D_H(z)/\rs$. For comparison we also plot the \shoes \citep{Riess:2021jrx} measurement corresponding to $H_0 = 73.04 \pm 1.04$ km/s/Mpc, although this is not used in any of the presented fits. It is offset from $z=0.0$ for visibility, and we divide by the fiducial \lcdm\ $\rs$ (solid purple star) or the \modrec\ best-fit $\rs$ (faint purple star). \textit{Bottom panel:} Similar to the other two panels, but for the angle-averaged distance $D_V(z) = (z D_A(z)^2 D_H(z))^{1/3}$. Only the solid data points are actually used in the DESI likelihood; the faded points are computed from the $D_A / \rs$ and $D_H / \rs$ measurements and are shown for illustrative purposes.}
\label{fig:distance} 
\end{figure}

Fig.~\ref{fig:distance} shows the DESI measurements of the three different distance scale variables: the transverse angular size of the BAO feature $D_A/\rs$, the line-of-sight BAO feature $D_H/\rs$, and the isotropic feature $D_V/\rs$. We also display the best-fit predictions for these variables as a function of redshift in the \lcdm, $w_0 w_a$CDM, and \modrec\ models. The top panel suggests that it is the transverse BAO measurement in the $z=0.71$ redshift bin which is primarily driving the trend seen in the \modrec\ best fit (red solid line), which returns to near the fiducial if that point is excluded (red dot-dashed line). 

The \modrec\ model does not alter the functional form of the redshift-distance relationship, so the differences seen in Fig.~\ref{fig:distance} are due to changes in the best-fitting $H_0$, $\omega_{\rm m}$, and $\rs$ relative to \lcdm, as discussed in LKC. The $w_0 w_a$CDM model (blue and black lines) does change this functional dependence, allowing for more complex shapes in these distance variables. The $w_0 w_a$CDM introduces a dip relative to the \lcdm\ best fit near $z=0.5$, where the DESI measures values of $D_V/\rs$ which are lower than what is predicted by, e.g., \planck\ assuming \lcdm. The result is a lower expansion rate today than any of the other models considered here, as is evident from the middle panel. We note that the combination of \planck+DESI+DESY5 (black line in Fig.~\ref{fig:distance}) favors $w_0 w_a$CDM over \lcdm\ at the $\sim 4 \sigma$ level \citep{DESI:2024mwx}.

The best fits in both the $w_0 w_a$CDM and \modrec\ models have comparable $\Delta \chi^2 \sim -7$ relative to \lcdm\ for the same data combination, however we note that the \modrec\ model has five more free parameters than does the $w_0 w_a$CDM model. 

The benefit of this extra freedom in the model is apparent when we consider the $\Delta \chi^2$ once \shoes\ data are included in the fit. To do so, we find the best fit to the \planck+DESI+\shoes data combination for both \lcdm\ and \modrec. For the \shoes\ measurement, we use a Gaussian likelihood given by $H_0=\ 73.04 \pm 1.04$ km/s/Mpc \citep{Riess:2021jrx}. We find that $\Delta \chi^2_{\rm Planck} = -11$, $\Delta \chi^2_{\rm DESI} = 3$, and $\Delta \chi^2_{\rm \shoes} = -15$, so that \modrec\ improves the quality of fit over \lcdm\ by $\Delta \chi^2_{\rm total} = -23$, a substantial payoff for the additional seven degrees of freedom. 

The increase of three in the DESI $\chi^2$ brings it to 15.8 for its 12 data points. Given the joint fit, it is not clear how to count the number of parameters DESI is constraining, but in the \modrec\ (and \lcdm) model space the most that can be is 2 (as we explain below) and the probability for $\chi^2$ to exceed $15.8$ for 12 - 2 = 10 degrees of freedom is 11\%. 

A more detailed Bayesian model comparison might further clarify the relative strengths of these models, but we have kept such a comparison outside the scope of this exploratory work.

\section{Discussion \label{sec:discussion}}

Here we discuss why the DESI data lead to an increased value of $H_0$ in the \modrec\ model, implications for the ``negative neutrino mass problem,'' the favored ionization history, and finally the tension with uncalibrated SN data. 

\subsection{Why do DESI data lead to a high $H_0$? \label{sec:H0_explained}} 

\begin{figure*}[t]
    \centering
    \includegraphics{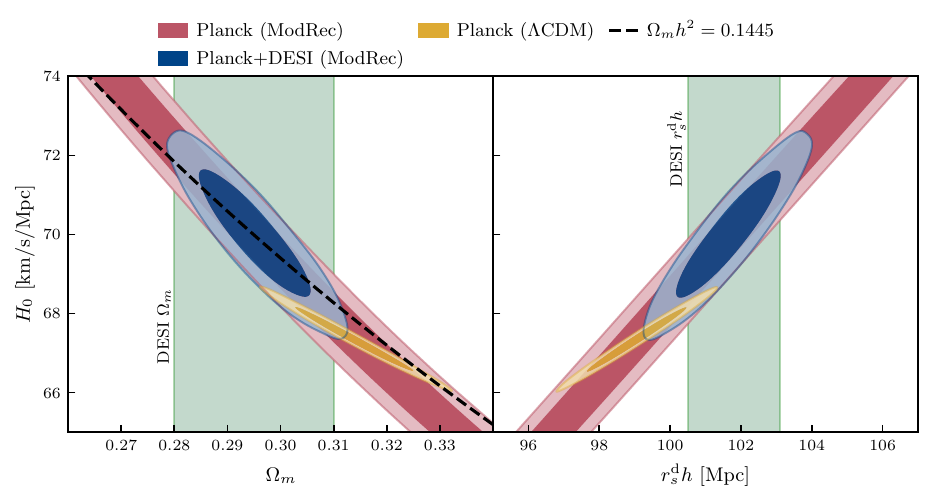}
    \caption{Parameter constraints in the $\Omega_{\rm m} - H_0$ and $\rs h - H_0$ planes. Green bands indicate 1D 68\% confidence intervals from DESI data when $\Omega_{\rm m}$ and $\rs h$ are treated as free parameters. The black dashed line corresponds to constant $\omega_{\rm m} \equiv \Omega_{\rm m} h^2 = 0.1445$, the mean value in the \planck\ (\modrec) chain.}
    \label{fig:H0-Omegam-rdh}
\end{figure*}

By themselves, under the assumption of \lcdm, BAO data are sensitive only to the combination $\rs h$ and $\Omega_{\rm m}$. When these are treated as free parameters, the DESI collaboration finds 68\% confidence intervals \citep{DESI:2024mwx}
\begin{align}\label{eqn:DESI_constraint}
    \Omega_{\rm m} &= 0.295 \pm 0.015\ {\rm and}\\
    \rs h &= (101.8 \pm 1.1)\ {\rm Mpc}.
\end{align}
We show these constraints as vertical bands in Fig.~\ref{fig:H0-Omegam-rdh}. Recall that the \modrec\ model, as we use it here, is \lcdm\ plus freedom in recombination. So given the \modrec\ model, the joint constraint on these two quantities is all the information in the DESI BAO data. We also show the constraints from \planck\ alone assuming \lcdm\ and then assuming the \modrec\ model. 

Before getting to the interaction of the \planck\ and DESI constraints, let us review how one can infer $H_0$ and $\Omega_{\rm m}$ from \planck\ data in \lcdm. The sensitivity of acoustic peak separations to $\theta_{\rm s}$ leads to a tight determination of $\theta_{\rm s}$, which in the \lcdm\ model with standard recombination leads to a strong constraint on $\Omega_{\rm m} h^3$ \cite{2dFGRSTeam:2002tzq,Planck:2018vyg,Lee:2022gzh}. The resonant driving of acoustic oscillations due to gravitational potential decay at horizon crossing, and the dependence of the amount of that decay on $\rho_{\rm m}/\rho_{\rm rad}$, leads to a tight constraint on the matter density $\Omega_{\rm m} h^2$ \cite{Hu:1996mn}. Combining these two constraints leads to constraints on $\Omega_{\rm m}$ and $H_0$, as can be seen from the \planck\ contours (assuming \lcdm) in Fig.~\ref{fig:H0-Omegam-rdh}.

The translation of the $\theta_{\rm s}$ constraint to a constraint on $\Omega_{\rm m} h^3$ relies on the determination of the sound horizon which, within \lcdm\, is possible using the standard recombination model. With the introduction of the \modrec\ freedom, that ability to infer the sound horizon is greatly weakened and the $\Omega_{\rm m}h^3$ constraint is effectively lost. In contrast, the interpretation of the radiation-driving effect (to deliver a constraint on $\Omega_{\rm m}h^2$) is not as reliant on the sound horizon determination and the one-dimensional constraint that remains in this plane is very close to being a constraint on the combination $\Omega_{\rm m} h^2$. One  can see this as the degeneracy direction is nearly along a contour of constant $\Omega_m h^2$. We show one corresponding to the \planck\ (\modrec) best-fit value of $\Omega_{\rm m}h^2 = 0.1445$. 

Inferences of the baryon density and matter density are all that are needed to determine $\rs$ in the \lcdm\ model \cite[e.g.][]{Knox:2019rjx}. With $H_0$ determined as well, the product $\rs h$ is of course also determined, as can be seen in the right panel of Fig.~\ref{fig:H0-Omegam-rdh}. But with the introduction of the \modrec\ freedom, $\rs h$ becomes unconstrained.

Either one of the two DESI BAO constraints on their own, added to the \planck\ constraints in the \modrec\ model, tighten the posterior for $H_0$ and shift it to a higher mode than one has given \planck\ and \lcdm. In fact they do so in a highly consistent manner with \planck+DESI $\Omega_{\rm m}$ resulting in $H_0 = 70.3 \pm 2.2$ km/s/Mpc, and \planck+DESI $\rs h$ resulting in $H_0 = 70.2 \pm 1.4$ km/s/Mpc. When using the full DESI data, we find $H_0 = 70.0 \pm 1.1$ km/s/Mpc. As one can see in these $H_0$ error bars and in the figures, the more $H_0$-informative of the two DESI constraints, is the $\rs h$ constraint. \citet{Pogosian:2024ykm} similarly noted that the higher value of $\rs h$ from the DESI data, when combined in their case with limited information from the \planck\ data, leads to a higher value of $H_0$. 

\subsection{Sum $\nu$s is better news}
In a recent paper \cite{Craig:2024tky} titled ``No $\nu$s is good news", the authors phenomenologically defined $\sum \tilde m_\nu$ such that at positive values it is identical to the sum of neutrino masses, and for negative values it artificially enhances CMB lensing power, by the same fractional amount as the deficit caused by $\sum m_\nu =-\sum \tilde m_{\nu}$ (see text). They then found that the combination of \planck, ACT DR6 lensing \cite{ACT:2023kun}, and DESI BAO data lead to a posterior probability distribution that peaks at $\sum \tilde m_\nu = -160$ meV with the minimal $\sum \tilde m_\nu = 58$ meV disfavored at 3$\sigma$. This is an interesting result suggesting that some element of the cosmological model used to interpret these datasets is wrong, or the errors in at least one of the datasets are underestimated.  

\citet{Craig:2024tky} also point out that the lensing power spectrum amplitude scales as
\begin{equation}
    C_L^{\kappa \kappa} \propto A_s (\Omega_{\rm m} h^2)^2 \left(1-0.02 \left[\frac{\sum m_\nu}{58\ {\rm meV}}\right]\right).
\end{equation}
Clearly the CMB lensing power spectrum is also sensitive to $n_{\rm s}$. We incorporate that by defining 
\begin{equation}
A_{\rm s}^\nu \equiv A_{\rm s} \times (0.1/0.05)^{n_{\rm s} - 1}
\end{equation}
since $k = 0.1$/Mpc is approximately where the combination of ACT CMB power spectrum measurement has the highest signal-to-noise ratio on neutrino mass \cite{Pan:2015bgi,ACT:2023kun}, and $A_s$ here (and in the \planck\ papers) is defined at $k=0.05/$Mpc.  
We note that, as can be seen in Table~\ref{tab:parameters}, the one-dimensional marginal posterior probability distribution of $A_s^\nu \omega_{\rm m}^2$ given the \modrec\ model and \planck+DESI BAO data, has a 3.3\% higher mode compared to the posterior obtained for \lcdm\ (with the same datasets).  We expect this upward shift would move the mode by about $0.033/0.02 \times 58$ meV $= 96$ meV toward positive $\sum \tilde m_\nu$, reducing the tension with the  minimal neutrino mass sum from $3\sigma$ to less than $2\sigma$. That calculation neglects an increase to $\sigma$ that we expect will occur as well, further reducing the tension. We note that the 95\% confidence upper limit on $A_{\rm s}^\nu \omega_{\rm m}^2$ inferred from \planck\ + DESI BAO increases by 5.5\% when switching from \lcdm\ to \modrec, which is even more than the shift in the mean, reflecting increased uncertainty.  

\subsection{Best-fit ionization history}

\begin{figure}[t!]
    \centering
    \includegraphics{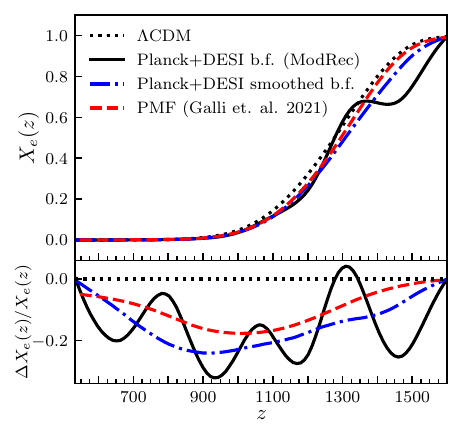}
    \caption{The best-fit ionization history from the \planck+DESI chain, a smoothed version of this, the standard \lcdm\ ionization history, as well as an ionization history from a PMF model following the treatment of \citep{Galli:2021mxk}, with clumping factor $b=0.5$. }
    \label{fig:rec_history} 
\end{figure}

In Fig.~\ref{fig:rec_history}, we present the ionization history that best fits the \planck+DESI data in the \modrec\ model. The most striking features of this ionization history are the large decrement in $\Xe$ at $z \gtrsim 1350$ as well as the oscillations in $\Delta \Xe / \Xe$, both of which outwardly seem unlikely to arise in a physical model. However, these features are not essential to provide a good fit to the data. To demonstrate this, we construct a smoothed version of the best-fit $\Delta \Xe(z) / \Xe(z)$ by interpolating between the approximate inflection points of the oscillatory $\Delta \Xe(z) / \Xe(z)$. This smoothed version is shown as the blue curve in Fig.~\ref{fig:rec_history}.

\begin{figure}[t]
    \centering
    \includegraphics{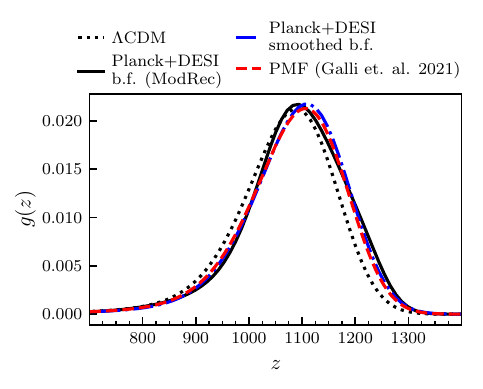}
    \caption{The visibility functions corresponding to the recombination histories presented in Fig.~\ref{fig:rec_history}. }
    \label{fig:visibility} 
\end{figure}

Compared to the oscillatory recombination history, the smoothed version has $\Delta \chi^2 = 14$ relative to \lcdm\ while keeping the cosmological parameters fixed. We then find the combination of cosmological parameters which best fit \planck+DESI given this ionization history data. The resulting best fit has $\Delta \chi^2 = 4.7$ relative to the unsmoothed version, and $\Delta \chi^2 = - 2$ relative to the \lcdm\ best fit. The middle column of Table~\ref{tab:modrec_comparison} shows parameter constraints when this is done. We conclude from this that while the oscillations do improve the fit by $\Delta \chi^2 \sim (\text{a few})$, they are not an essential feature of the \modrec\ solution: the smoothed $\Delta \Xe(z) / \Xe(z)$ deviates from the standard ionization scenario by about 25\% near the peak of visibility (see Fig.~\ref{fig:visibility} and discussion below), and yet fits the data as well as the standard scenario.

For comparison, we also show the ionization history given by a model with primordial magnetic fields, following the treatment of \citet{Galli:2021mxk}. The smoothed version of the \modrec\ best-fit $\Xe(z)$ is qualitatively similar to this recombination history, albeit with a larger overall deviation. 

\begin{table}
\caption{\label{tab:viz_moments} Moments of the visibility functions presented in Fig.~\ref{fig:visibility}. Moments higher than the mean are computed as central moments about the mean.}
\begin{ruledtabular}
\begin{tabular}{l c c c c}
Model                   & Mean        & Variance   & Skewness & Kurtosis \\
\hline
\lcdm                   & $1064$ & $9607$  & $-8.627 \times 10^{5}$ & $4.713 \times 10^{8}$\\
\modrec                 & $1086$ & $10064$ & $-8.976 \times 10^{5}$ & $5.177 \times 10^{8}$\\
\modrec\ (smooth)       & $1085$ & $9930$  & $-9.210 \times 10^{5}$ & $5.167 \times 10^{8}$\\
PMF                     & $1079$ & $10912$ & $-1.535 \times 10^{6}$ & $9.775 \times 10^{8}$\\
\end{tabular}
\end{ruledtabular}
\end{table}

We note that the \modrec\ solution, obtained in a model-independent and data-driven way, is indicative of accelerated recombination without specifying the underlying physical mechanism causing this. In Fig.~\ref{fig:visibility}, we show the visibility functions $g(z)$ corresponding to these ionization histories, and in Table~\ref{tab:viz_moments} we compile the first few moments of these visibility functions. As expected, the mean of $g(z)$ in each of the non-\lcdm\ models is shifted to a higher redshift, indicative of an earlier recombination. In addition, we note the general agreement between the moments of the \modrec\ and smoothed \modrec\ visibility functions, despite the rather large differences in $\Xe(z)$ for these two recombination histories and differing visibility functions. These moments provide a quantitative summary of the visibility functions in these models, and may be useful for comparisons to physical models.

\subsection{Tension in $\Omega_{\rm m}$ measurements}

Measurements of the matter density $\Omega_{\rm m}$ from uncalibrated SNe prefer higher values than what is obtained using CMB or BAO data alone. This disagreement is worsened within the \modrec\ model, with which we infer $\Omega_{\rm m} = 0.295 \pm 0.007$ using a combination of \planck\ and DESI data, a value which is slightly lower than what is obtained when assuming \lcdm. Under the \modrec\ (\lcdm) model, the significance of this disagreement is $1.9\sigma$ ($1.3\sigma$), $2.3\sigma$ ($1.9\sigma$), $3.15\sigma$ ($2.6\sigma$) when compared to the Pantheon+ \citep{Brout:2022vxf}, Union3 \citep{Rubin:2023ovl}, and DESY5 \citep{DES:2024tys} datasets, respectively. Table~\ref{tab:parameters} shows the inferred values of $\Omega_{\rm m}$ and the resulting tension with DESY5 for different models and data combinations.

The uncalibrated SN measurements are a challenge for models which restore concordance by modifying recombination. Unlike BAO data, the interpretation of the supernova measurements is unaffected by the recombination model. These data provide low-redshift distance measurements which constrain $\Omega_{\rm m}$ independently to CMB and BAO data. The $\Omega_{\rm m}$ inferences generally come out somewhat higher (e.g. DES find $\Omega_{\rm m} = 0.352 \pm 0.017$). Just as the DESI low $\Omega_{\rm m}$ inference leads, in combination with \planck\ under the assumption of \modrec, to earlier recombination and higher $H_0$, the combination of SN data with \planck\ assuming \modrec\ would lead to delayed recombination relative to that in \lcdm\ and a lower $H_0$, exacerbating the tension with SH0ES. We thus see that in combination with CMB data, supernova data can have an impact on the reconstruction of the ionization history. 

We can also consider what we would get from combining \planck, DES SNe, and DESI assuming the \modrec\ model. Note that taking the minimum variance average of $\Omega_{\rm m} = 0.352 \pm 0.017$ from DESI and $\Omega_{\rm m} = 0.295 \pm 0.015$ from DES we get $\Omega_{\rm m} = 0.320 \pm 0.011$. This is 1.5\% less than the \planck\ value of $\Omega_{\rm m} = 0.3153 \pm 0.0073$, assuming \lcdm. Combining this with the \planck\ \modrec\ value of $\Omega_{\rm m} h^2 = 0.144 \pm 0.005$ (increased by 1\% from the \lcdm\ value) we would get an $H_0$ value about 1\% higher than in \planck\ \lcdm. Including the DESI $r_s^d h$ information would boost this up further. Thus we expect that adding DES SNe to the \planck\ and DESI BAO datasets in the \modrec\ model would significantly moderate the preference for earlier recombination and higher $H_0$, but not completely eliminate or reverse it. 

\section{\label{sec:conclusion}Conclusion}

We have analyzed recent DESI BAO data, in combination with \planck\ CMB data, in the context of a modified recombination epoch. We find a preference for $H_0 = 70.0 \pm 1.1$ km/s/Mpc while retaining the cosmological constant as an explanation for dark energy. Additionally, we potentially point the way toward resolving the cosmological neutrino mass problem highlighted in \citep{Craig:2024tky}.

Our phenomenological framework allows us to identify ionization histories which explain the observations well in a model-independent and data-driven manner. When we do so here, we find that a smoothed version of the ionization history that best fits the \planck+DESI data bears qualitative similarities to recombination models in the presence of small-scale clumping. We find this notable in the context of the growing number of hints that \lcdm\ at small scales may require modifications. It is interesting that, in the context of the \modrec\ model used here, this seems to be the direction that DESI BAO data are now pointing as well.

There are also arguments that disfavor this scenario. Chief among these is the empirical success, with CMB data, of the \lcdm\ prediction for the ionization history. This success is particularly noteworthy given that the electron-proton-hydrogen atom system is significantly out of chemical equilibrium \citep{Zeld1968Rec, 1968ApJ...153....1P, Seager:1999km, Chluba:2010ca,2011PhRvD..83d3513A}, resulting in the dependence of CMB temperature and polarization power spectra on the ratio of atomic reaction rates relative to the expansion rate  \citep{Zahn:2002rr, Ge:2022qws}. Second, our results are strongly driven by the DESI $z=0.71$ BAO measurement which uses luminous red galaxies as tracers for structure. The DESI collaboration note that DESI and SDSS data are discrepant at the $\sim 3\sigma$ level within this redshift bin, with the current best explanation for this discrepancy being a sample variance fluctuation. Whether this discrepancy persists as more data are collected and analyzed will be important for interpreting what DESI data imply about the recombination model. Finally, the modified recombination scenario analyzed here increases the tension in $\Omega_{\rm m}$ somewhat, by 0.4 to 0.5 $\sigma$ depending on SN dataset.

As discussed in LKC and in \cite{Prabhu:2024qix} forthcoming small-scale CMB temperature and polarization data will have the sensitivity required to constrain the scenario considered here. Whether DESI data are pointing toward something surprising during recombination is thus a question we expect to be answered in the near future.

Confirmation of modified recombination would also require identification of a physical mechanism which can deliver an ionization history consistent with what we have reconstructed from the data. The search for such a model might be aided by comparison of the first few moments of the visibility function, for which our analysis provides clear targets. The reconstruction process itself could also be improved. As we have seen, some of the features obtained with \modrec\ are not unique (or required); an improved parameterization might avoid such features. With the advent of more CMB anisotropy and BAO data we can look forward to exploring these questions.

{\small
\section*{Acknowledgements}
The authors thank Levon Pogosian for useful comments and discussion.
LK and GL were supported in part by DOE Office of Science award DESC0009999. LK also thanks Michael and Ester Vaida for their support via the Michael and Ester Vaida Endowed Chair in Cosmology and Astrophysics.
JC was supported by the ERC Consolidator Grant {\it CMBSPEC} (No.~725456) and by the Royal Society as a Royal Society University Research Fellow at the University of Manchester, UK (No.~URF/R/191023).}

\bibliography{apssamp}

\clearpage

\end{document}